\documentclass{article}

\usepackage{arxiv}

\usepackage[utf8]{inputenc} 
\usepackage[T1]{fontenc}    
\usepackage{hyperref}       
\usepackage{url}            
\usepackage{booktabs}       
\usepackage{amsfonts}       
\usepackage{nicefrac}       
\usepackage{microtype}      
\usepackage{lipsum}		
\usepackage{graphicx}
\usepackage{natbib}
\usepackage{doi}
\usepackage{amsmath}

\title{Enhancing Deep Learning Model Explainability in Brain Tumor Datasets using Post-Heuristic Approaches}

\author{Konstantinos Pasvantis\\
	Department of Applied Computer Science\\
	University of Macedonia\\
	Thessaloniki, Greece, 54636 \\
	\texttt{aid23005@uom.edu.gr} \\
	\And
	Eftychios Protopapadakis \\
	Department of Applied Computer Science\\
	University of Macedonia\\
	Thessaloniki, Greece, 54636\\
	\texttt{eftprot@uom.edu.gr} }

\renewcommand{\shorttitle}{\textit{arXiv} Template}

\begin{document}
\maketitle

\begin{abstract}
The application of deep learning models in medical diagnosis has showcased considerable efficacy in recent years. Nevertheless, a notable limitation involves the inherent lack of explainability during decision-making processes. This study addresses such a constraint, by enhancing the interpretability robustness.  The primary focus is directed towards refining the explanations generated by the LIME Library and LIME image explainer. This is achieved throuhg post-processing mechanisms, based on scenario-specific rules. Multiple experiments have been conducted using publicly accessible datasets related to brain tumor detection. Our proposed post-heuristic approach demonstrates significant advancements, yielding more robust and concrete results, in the context of medical diagnosis.
\end{abstract}


\section{Introduction}
Deep learning has become a powerful tool with many uses in the last few years, especially in the medical domain. Its capacity to evaluate complicated data, such as medical imaging, has demonstrated great potential for prognostic and diagnostic purposes (\cite{tran2021deep}, \cite{zhu2020application}, \cite{katsamenis2020transfer}, \cite{huang2021accurate}, \cite{voulodimos2018deep}, \cite{voulodimos2021deep}). Deep learning's extensive capabilities have made progress in medical image analysis possible, allowing for the more accurate and efficient diagnosis of a variety of illnesses.

However, the deployment of deep learning models in medical image analysis is not without challenges (\cite{altaf2019going}, \cite{razzak2018deep}). One major issue with these models is their lack of explainability. Because deep neural networks are complex, their decision-making processes are frequently transparent, which makes it difficult for medical experts to understand and accept the outcomes. This problem is particularly important for medical applications because misinterpretation might have serious consequences.

Researchers have looked into a number of ways to improve the interpretability of deep learning models in order to close the explainability gap, especially when it comes to medical imagery (\cite{uzunova2019interpretable}, \cite{zhang2017mdnet}, \cite{dravid2022medxgan}). Such approaches could be: a) Model-specific methods, e.g. saliency maps or activation maximization (\cite{katzmann2021explaining}, \cite{ann2021multi}) or b) Model-agnostic methods, e.g. partial dependence plots or surrogate models (\cite{kim2021model}, \cite{grassucci2023grouse}, \cite{yang2020dies}, \cite{peng2021explainable}).  Even though progress has been achieved in this area, there is always opportunity for enhancement, particularly when it comes to improving the results' interpretability for certain applications.

This paper explores the relationship between medical image analysis and deep learning, with a particular emphasis on applying explainable AI to detect brain tumors from MRI-images. Understanding the difficulties in obtaining results that are transparent, we use an explainability method that is specific to the complexities of medical image analysis. Most importantly, we contribute to a later improvement of this method with the goal of offering more accurate and useful information on the existence of brain tumors.

\section{Related Work}
The efficiency and accuracy of medical image analysis has greatly increased in recent years due to the impressive progress made in the application of deep learning algorithms. These advances have mostly focused on segmentation and classification, which are critical roles in medical image processing. Diverse deep learning structures have been investigated to address the difficulties present in these challenges.

Deep learning plays a crucial role regarding image classification in medical field too. CNNs, such as AlexNet \cite{krizhevsky2012imagenet} and VGGNet \cite{DBLP:journals/corr/SimonyanZ14a}, have been influential in classifying medical images into distinct categories. Transfer learning methods have also made it easier to apply previously learned CNN models to medical image classification tasks, which has enhanced diagnostic capabilities. AlexNet, for example, has shown success in classifying skin lesions in dermatological images, contributing to the early identification of skin cancers \cite{hosny2020classification}. Another notable example of transfer learning is the adaptation of a pre-trained ResNet model for the detecting COVID-19 in chest X-rays \cite{hamlili2022transfer}.

In the field of neuroscience,in order to mitigate the serious health risks that brain tumors offer, fast and accurate brain tumor detection is essential. Early detection has a major impact on treatment outcomes in addition to allowing timely intervention. In this context, deep learning approaches have transformed the processing of brain tumor images. In particular, architectures like U-Net have proven useful in accurately identifying brain tumors from MRI scans, supplying vital data for treatment planning (\cite{ilhan2022brain}, \cite{allah2023edge}). Moreover, the categorization of brain images has been improved with the use of  custom CNN architectures and transfer learning methods.

As deep learning techniques are still transforming medical image processing, there is an increasing need for these complicated models' decisions to be transparent and understandable. Explainability, also known as the interpretability of artificial intelligence (AI) systems, is becoming more and more important, especially in healthcare applications where it is critical to comprehend the reasoning behind a diagnosis (\cite{tjoa2020survey}, \cite{de2023explainable}). Explainable AI (XAI) aims to provide information about how and why specific conclusions are made from complicated datasets, simplifying the decision-making processes of deep learning models in the larger context of image processing.

Explainability becomes a critical element in the field of medical image analysis, ensuring that healthcare professionals can trust and confidently understand the insights offered by AI models in the diagnosis and treatment planning of a variety of illnesses, including brain tumors (\cite{zeineldin2022explainability}, \cite{ahmed2023identification}).

Regarding brain tumor detection from images, there are many methods used for explainability, gradient-based or perturbation-based \cite{zeineldin2022explainability}. Local Interpretable Model-agnostic Explanations
(LIME)\cite{ribeiro2016should}, is a perturbation-based method that finds the segments contributing to a model's prediction for a given image. There are already researches focused in explaining decisions made from pre-trained deep learning models (\cite{gaur2022explanation}, \cite{haque2024neuronet19}), using this method, but this method alone may not produce a proper explanation to make things better for the user. 

\subsection{Research Challenges}
\label{subsubsec:res_challenges}

One significant challenge posed by the LIME library is the inherent difficulty of producing segments that correspond with the semantic interpretation of the image \cite{hryniewska2022limecraft}. As a result, a segment that has no information can appear as important as other segments, missleading the user that is trying to understand the prediction. 

Furthermore, LIME is sensitive to minor changes in the input image.  The stability of LIME's explanations can be significantly affected by factors such as the addition of noise posing the possibility of inconsistent results.

\subsection{Our Contribution}
\label{subsec:contrib}

Based on the existing bibliography and at the best of our knowledge, at this point we have no end to end architectures specifically designed for refining explainability results through image post-processing techniques. To address this gap, our study provides a new refining method focused toward enhancing the interpretability of image-based explanations. By integrating this refinement step into the explainability workflow, we aim to provide a more comprehensive and reliable solution for addressing the challenges posed by existing frameworks, especially in the domain of medical image analysis.

\section{Proposed Methodology}
Let $I \in\mathbb{Z}^{w \times h }$ a grayscale image, originating from an MRI scanner, and $ t \in \{0,1\}$ the decision of a deep learning model regarding the existence or not of a tumor.  Then, a model-agnostic technique (sec. \ref{subsec:explainability_techniques}) generates a heatmap, $H \in\mathbb{R}^{w \times h }$, over image $I$. The heatmap indicates regions of interest, over $I$, which have contributed to generating an output $t$. Ideally, the most prominent regions should include a high portion of the tumor area. 

\begin{figure}[h]
\centering
    \includegraphics[width=12cm]{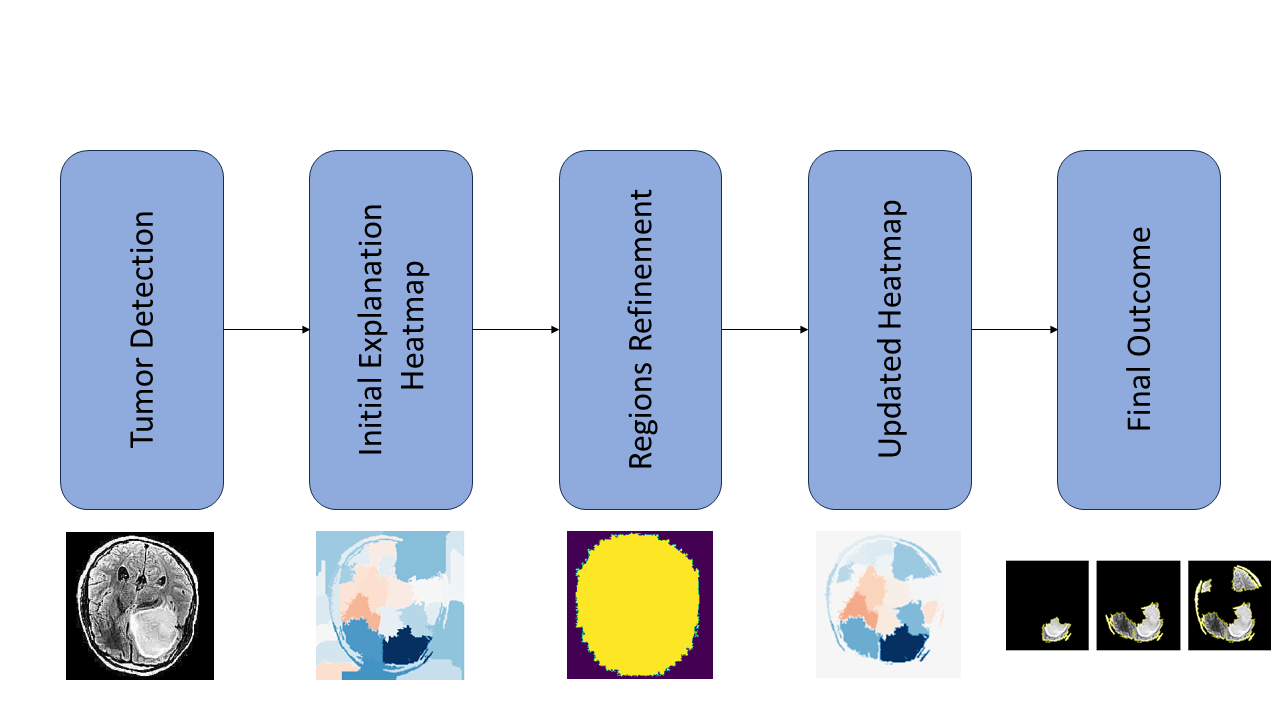}
    \caption{Proposed Methodology}
    \label{fig:prop_meth}

\end{figure}

The adopted approach introduces an additional refinement mechanism (sec. \ref{subsec:refinements}), $R(I,H)$, which considers both $I$ and $H$, and eliminates non-informative segments of $H$, based on a combination of image morphology operations and post-processing heuristics, so that $R(I,H) \rightarrow H^{(R)}$. $H^{(R)}$ is the refined version of $H$, retaining the most appropriate segments, related to brain and tumor geometry. Figure \ref{fig:prop_meth} demonstrates the process.

\subsection{Employed Deep Learning Architectures}
\label{subsec:deep_learn_architectures}
In this work we handle the brain tumor detection as a binary classification problem, using as input a gray scale image, say $I$. We try to establish a prediction model, $f(x)\rightarrow t$, $ t \in \{0,1\}$, so that given the image $I$, 
$$f(I)=\begin{cases}
			0, & \text{if $I$ has not brain tumor}\\
            1, & \text{if $I$ has brain tumor}
		 \end{cases}
$$
The process incorporated the paradigm of transfer learning, since it can provide significant advantages in medical applications \cite{maganaris2022evaluating}. In particular, three Deep Learning pre-trained models were used, InceptionV3 \cite{szegedy2016rethinking}, ResNet50V2 \cite{he2016identity} and NasNetLarge \cite{zoph2018learning}. 

In our study, the pretrained layers of each model were frozen to retain their weights from the Imagenet Dataset, and a custom head is added in order to adapt to our dataset. We use the same parameters for each of these pretrained models, and a head with global average pooling and dense layers concluding with a softmax layer. Training involves utilizing the adam optimizer, and categorical crossentropy loss, over 10 epochs. An early stopping criterion is implemented in order to reduce overfitting.  

\subsection{Model's Explainability}
\label{subsec:explainability_techniques}

The Lime Image Explainer (LIE), from LIME library, is employed for creating an explanation, given an images $I_c$, for which $t_c = 1$, i.e. positive to cancer.

At first, given the image $I_c$, LIE creates a new set of images, say $Z = \{I_c^{(1)}, ..., I_c^{(n)}\}$, with the same dimensions $w \times h$. The process can be summarized as follows: a) A segmentation algorithm, e.g. QuickShift, operates over image $I_c$, generating $d$ segmented areas. b) Then, a new image instance, $I_c^{(k)}$ is created, by maintaining a random number, $m<d$, of the original segments. c) Repeat the process until a predefined number of images are generated. 

The proposed approach generates multiple copies of the image $I_c$, with missing areas, corresponding to some of the $d$ segments. Theoretically, such a process may generate up to $n= \binom{d}{0} + \binom{d}{1} + \dots + \binom{d}{d}$ new image instances.

We use the prediction model, described in section \ref{subsec:deep_learn_architectures} to predict the outcomes, using $Z$ as inputs. Each instance $I_c^{(k)} \in Z$ is fed to the deep learning model and the corresponding output, say $t^{(k)}$ is generated. LIE calculates the weights corresponding to each segment area, creating a sparse linear model, which approximates the outputs of the deep learning model $f(x)$ \cite{rabold2020enriching}. That way, the weights highlight the importance of each segment in contributing to the model's decision, providing a local and interpretable understanding of the black-box model's behavior within a specific region of the input space.

The result is a heatmap, of the form: $H = \{S_i: Importance_i, \dots, S_j: Importance_j\}$, where $S_i$ represents the ID of the segment, and $Importance_i$ denotes the corresponding importance value or the weight assigned to that segment by LIE. Our interest is identifying the best $n$ areas from the heatmap $H$ provided by the explanation.

\begin{figure}[h]
\centering
    \includegraphics[scale=0.6]{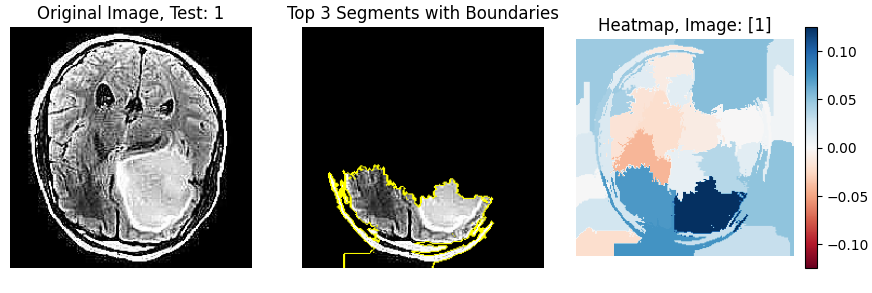}
    \caption{Top segments contributing to the prediction, with the produced heatmap from LimeImageExplainer}
\end{figure}

\subsection{Post-processing refinement mechanisms}
\label{subsec:refinements}

After examining the heatmap provided for the images, it turns out that certain segments are outside the bounds of the brain, having no meaningfull contribution, from a medical expert perspective, in the generated explanation. As such, multiple image operators, i.e. filters, could be considered to improve and refine the explanations' interpretability.

The core idea lies in the successful detection of the brain area, given an image $I$. In this case, the problem at hand could be addressed as an edge detection problem. There are multiple explicit approaches for the identification of edges in an image like Laplace, Canny, and Sobel, or implicit ones like Otsu and Li. 

Subsequently, detected edges are used to create a binary brain mask, which is equal in size to the original image. Each pixel in this mask is assigned a value of 1 if it belongs to the brain region and 0 otherwise. The generation of this brain mask is made easy by extracting the largest contour associated with the detected edges from each of the above techniques. 
The resulting brain mask, denoted as $BrM$, serves as a crucial element in our refinement process, providing a clear binary definition of the brain's spatial extent.

After the production of the brain mask for an image, our refinement mechanism introduces a criterion for retaining segments produced by LimeImageExplainer, for the same image. The LimeImageExplainer, produced a heatmap $H$ (sec. \ref{subsec:explainability_techniques}. We then retain only the segments if a substantial proportion, specifically 80\% or more, of their pixels fall within the pixels of the brain mask, and manually set the Importance of the other segments to 0.
$$Selected Segments = \{S_i | Pixels(S_i\cap BrM) \geq 0.8\cdot Pixels(S_i)\}$$
The result is a new heatmap $H'$, meaning a new dictionary that is: 
$$H' = \{S_i: Importance_i', \dots, S_j: Importance_j' \}, $$
where this time, the $Importance_i'$ is the same with the $Importance_i$ if $Segment_i$ was retained, and 0 otherwise. This refined heatmap $H'$ provides a more accurate representation of the segments contributing to the model's prediction. 

To put it simply, the entire post-processing mechanism can be briefly represented as the refined explanation for an image ($Refined(I)$). This refined explanation is obtained from LIE, with the additional step of retaining segments based on the brain mask criteria. The integration of edge detection, brain mask generation, and segment retention than concludes in a post-refinement process, enhancing the clarity of the generated explanations.

\section{Experimental Setup}
\label{sec:exp_set}

All of the experiments conducted were employed using python, utilizing public libraries (Tensorflow, Skimage, Sklearn, Shapely, Lime, Matplotlib). The computations were performed in Google Colab with GPU acceleration enabled.

\subsection{Utilized Dataset}
\label{subsec:dataset}

The dataset utilized in this study is the \href{https://www.kaggle.com/datasets/preetviradiya/brian-tumor-dataset}{"Brain Tumor Dataset"} obtained from Kaggle. This publicly available dataset consists of 4602 MRI images capturing random instances of the brain. The images are categorized based on the presence or absence of a brain tumor, offering a wide range of samples for testing and training. The dataset includes various perspectives like axial, coronal, and sagittal views. To maintain consistency, only JPEG images in grayscale format were retained for analysis, ensuring a standardized input for the models and enhancing the reliability of the study's outcomes.

\subsection{Dataset Preprocessing}
\label{subsec:data_preprocessing}

Prior to training the models, the dataset underwent preprocessing steps to ensure consistency and suitability. Images were resized to 224$\times$224 pixels and normalized to standardize pixel values within the range of 0 to 1. To improve the dataset's quality and diversity, duplicate images were removed, resulting to a total of 4015 images. Furthermore, a Stratified K-Fold validation strategy with 5 splits was used. This methodology ensures a robust evaluation of the deep learning models' performance by guaranteeing that each fold retains the same distribution of classes as the original dataset. 

\subsection{Performance Metrics}
\label{subsec:perf_metrics}

Each deep learning model's performance was evaluated using typical metrics like accuracy, precision, recall and F1-Score. These metrics rely on key elements such as True Positives (TP), True Negatives (TN), False Positives (FP), and False Negatives (FN). This metrics are defined as: 
$$ Accuracy = \frac{TP + TN}{TP + TN + FP + FN}$$
$$ Precision = \frac{TP}{TP + FP}$$
$$ Recall = \frac{ TP}{TP + FN}$$
$$ F1 Score = 2 \cdot \frac{Precision\cdot Recall}{Precision + Recall}$$
In addition to these metrics, we introduce a new metric to evaluate how well the segments from LimeImageExplainer match the presence of tumors. Using the VGG Image Annotator, manual annotations of the tumor were performed on 271 images out of the 471 infected instances in the test set. These images were selected for their relative ease of annotation. This custom metric aims to quantify the percentage of the brain tumor included in the explanation's segments and we will refer to it as "Tumor Segment Coverage".

To be more specific, after the use of the VGG Image Annotator, a new mask that represents the location of the tumor is created, call it $Tum$. Meaning that a pixel of the original image $(x,y)$ belongs in the Tumor Mask, if-f this pixel is inside of the tumor polygon that is created. With this mask, we calculate the percentage of pixels that are both part of the mask, and the tumor annotation. 

In addition we will use another metric called "Brain mask Segment Coverage", and it will be the percentage of brain mask covered by the explanation's segments. Meaning that, after we produce the refined explanation, we calculate the percentage of the pixels that are part of the explanation, and also belongs in the brain mask that is created using an edge detector, as explained in section \ref{subsec:refinements}.

\subsection{Experimental Results}
\label{sec:exp_res}

The graphs below show the value of performance metrics for each model's predictions for every fold, with the last column presenting the summary score for  each model across all images.

\noindent
\begin{figure}[h]
    \centering
    \includegraphics[width=0.4\textwidth]{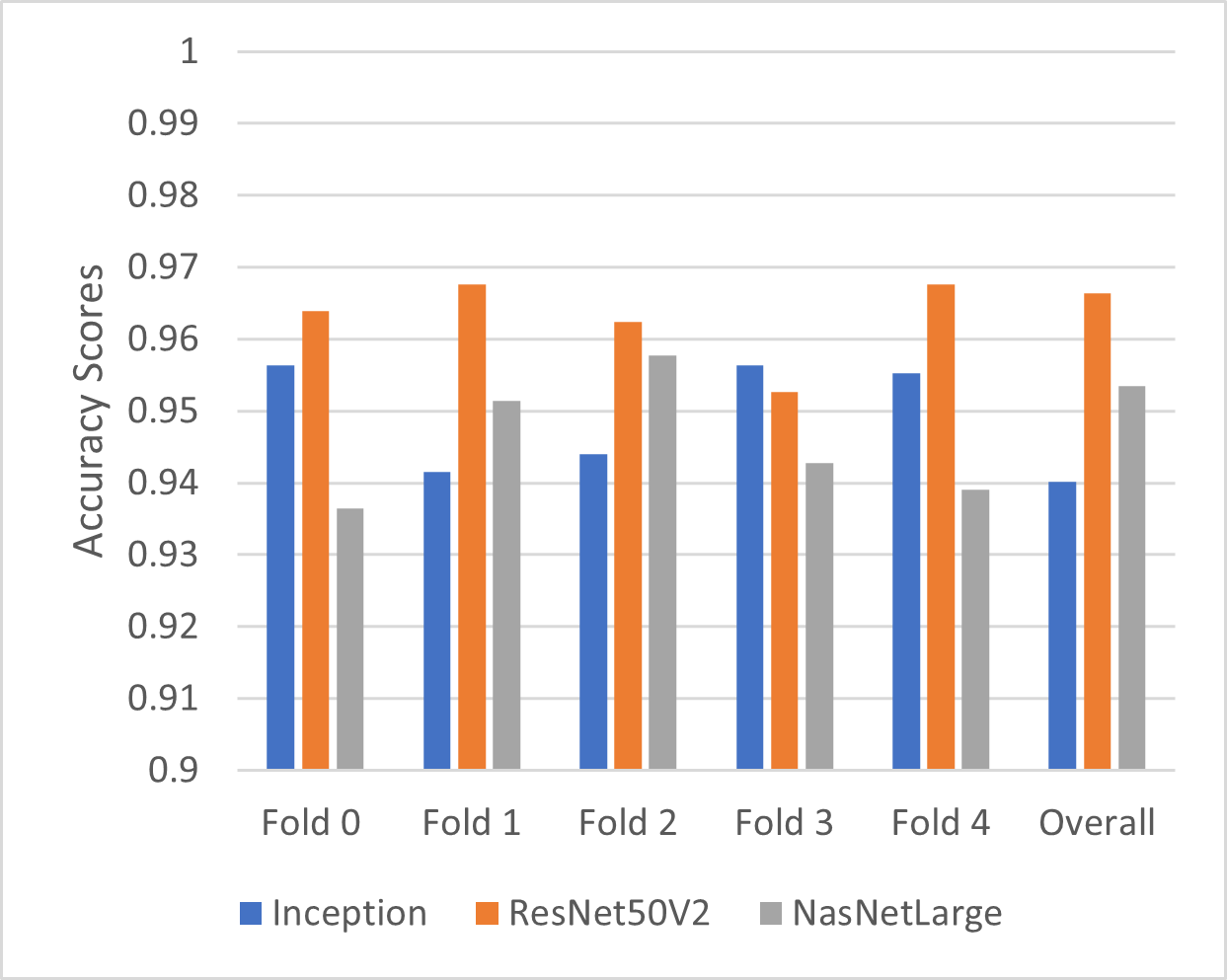}%
\includegraphics[width=0.4\textwidth]{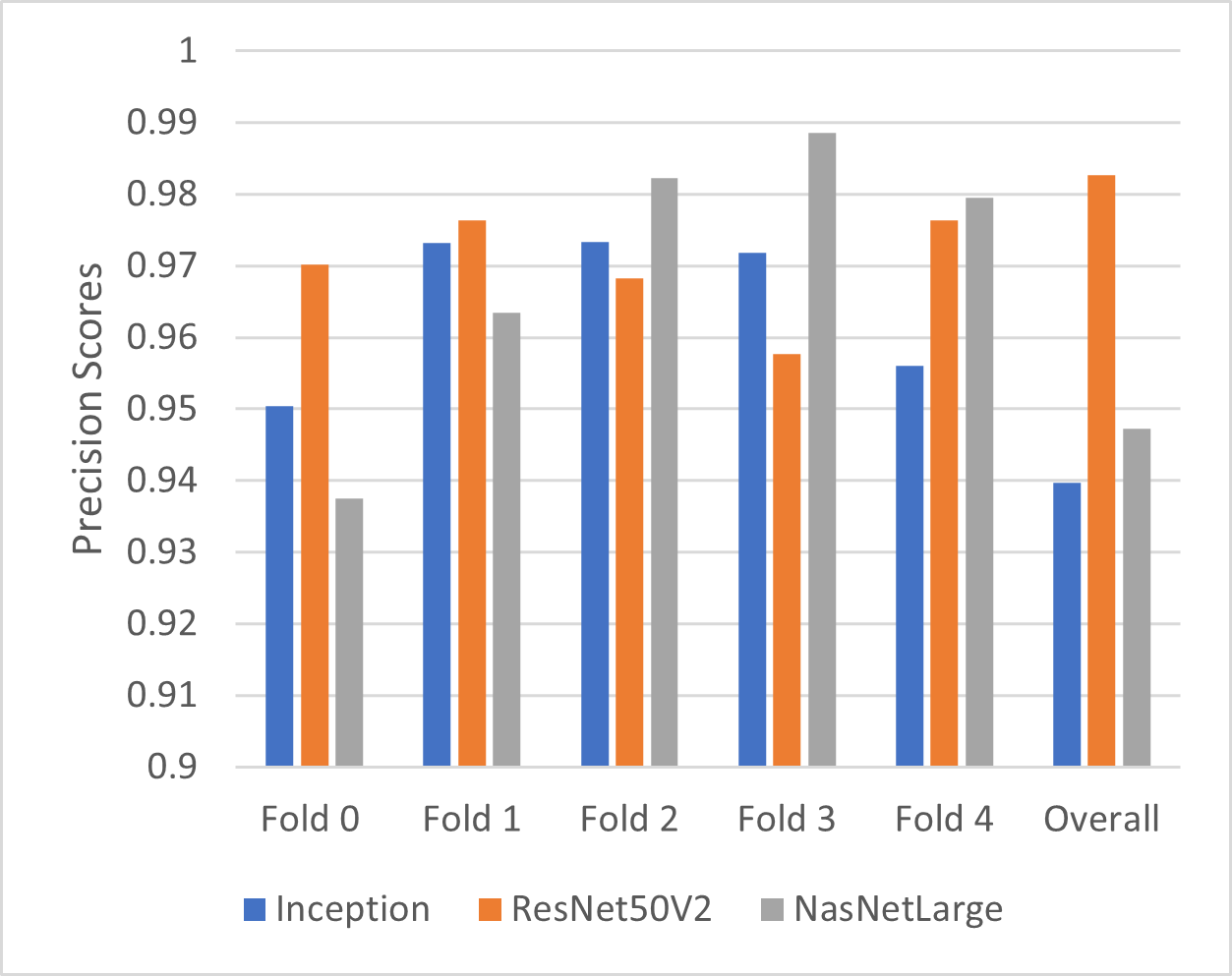}\\
\includegraphics[width=0.4\textwidth]{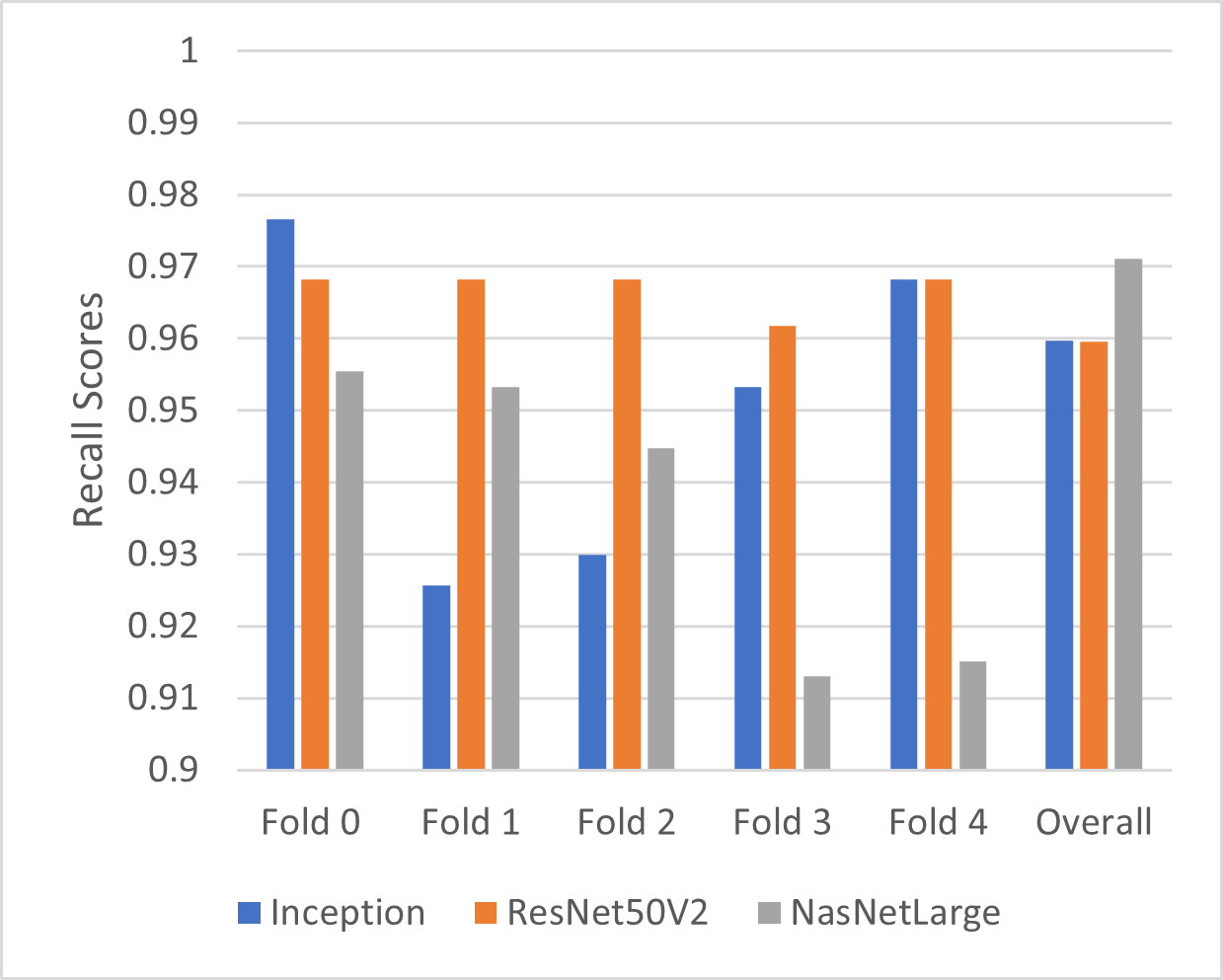}%
\includegraphics[width=0.4\textwidth]{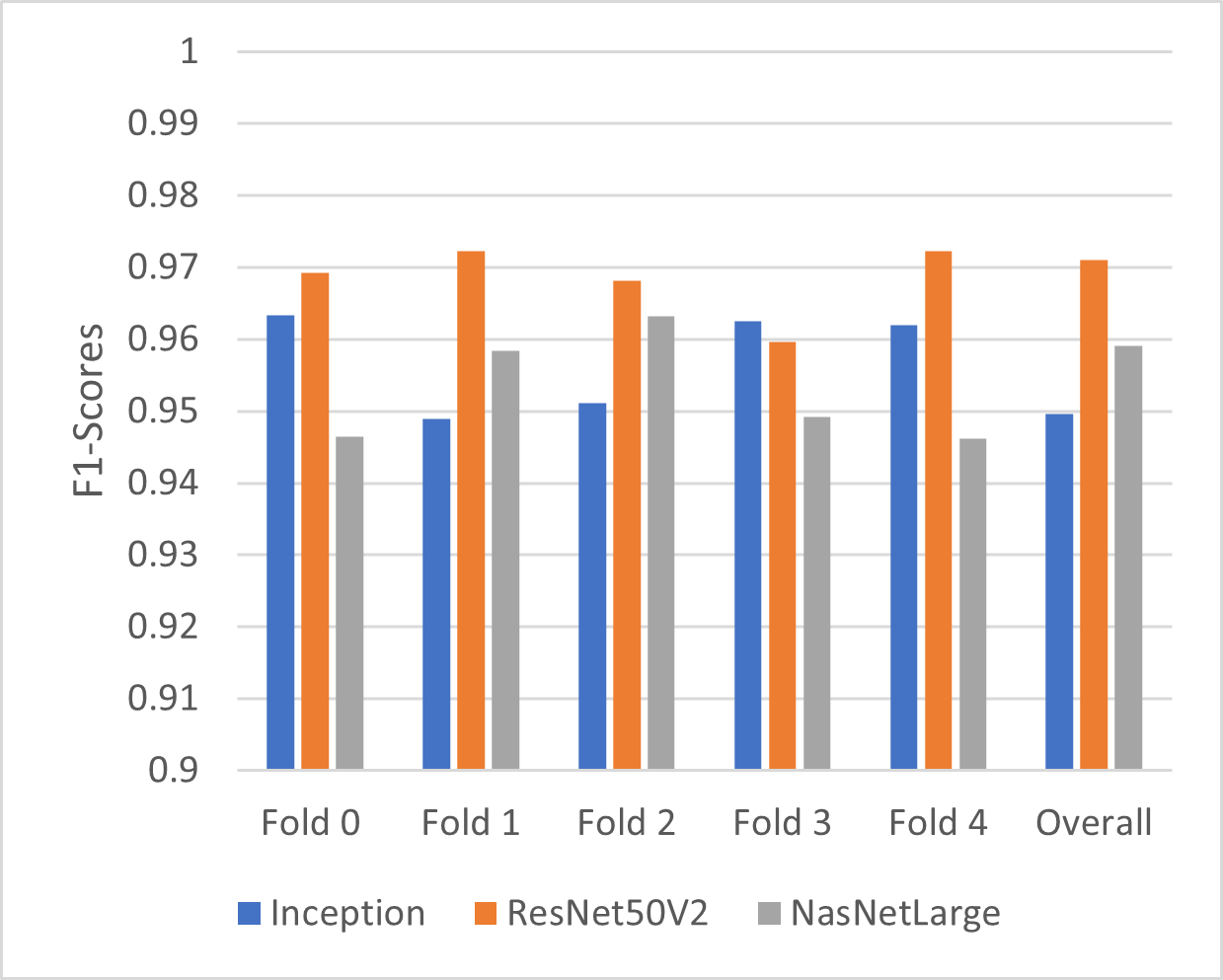}\par

    \caption{Performance Metrics of each model for each fold. The last collumn corresponds to each model after k-fold validation}

\end{figure}

To compare the pretrained models, we conducted statistical Mann-Whitney test on their F1 scores, considering both the scores obtained during cross-validation and after. The test showed that ResNet50v2 outperformed both InceptionV3 (p = 0.02) and NasNetLarge (p = 0.008). However, there was no significant difference between InceptionV3 and NasNetLarge (p = 0.39). Given these results, we chose ResNet50v2 for further analysis in the explainability part. 

As mentioned before, LIE was employed to provide explanations for model predictions before and after the proposed refinement. Without refinement, the explanations showed a suboptimal tumor segment coverage, averaging only 32.41\%. Following the introduction of the refinement mechanism, three different techniques (Canny, Laplace, Otsu's thresholding) were used for the production of the brain mask. To determine the best number of segments for generating meaningful explanations, we explored the impact of selecting the best 1, 3 and 5 segments using the refined LIME Image Explainer. Examining the tumor segment coverage, we found that relying on a single segment yielded an average coverage of 27.63\%, closely resembling the performance of picking the best 3 segments without our refinement. A substantial improvement was observed, with the average increasing to 50.28\% when we pick the best 3 segments. Employing 5 segments resulted in tumor coverage average of 63.84\%. 

\begin{figure}[h]
    \centering
    \includegraphics[width=10cm]{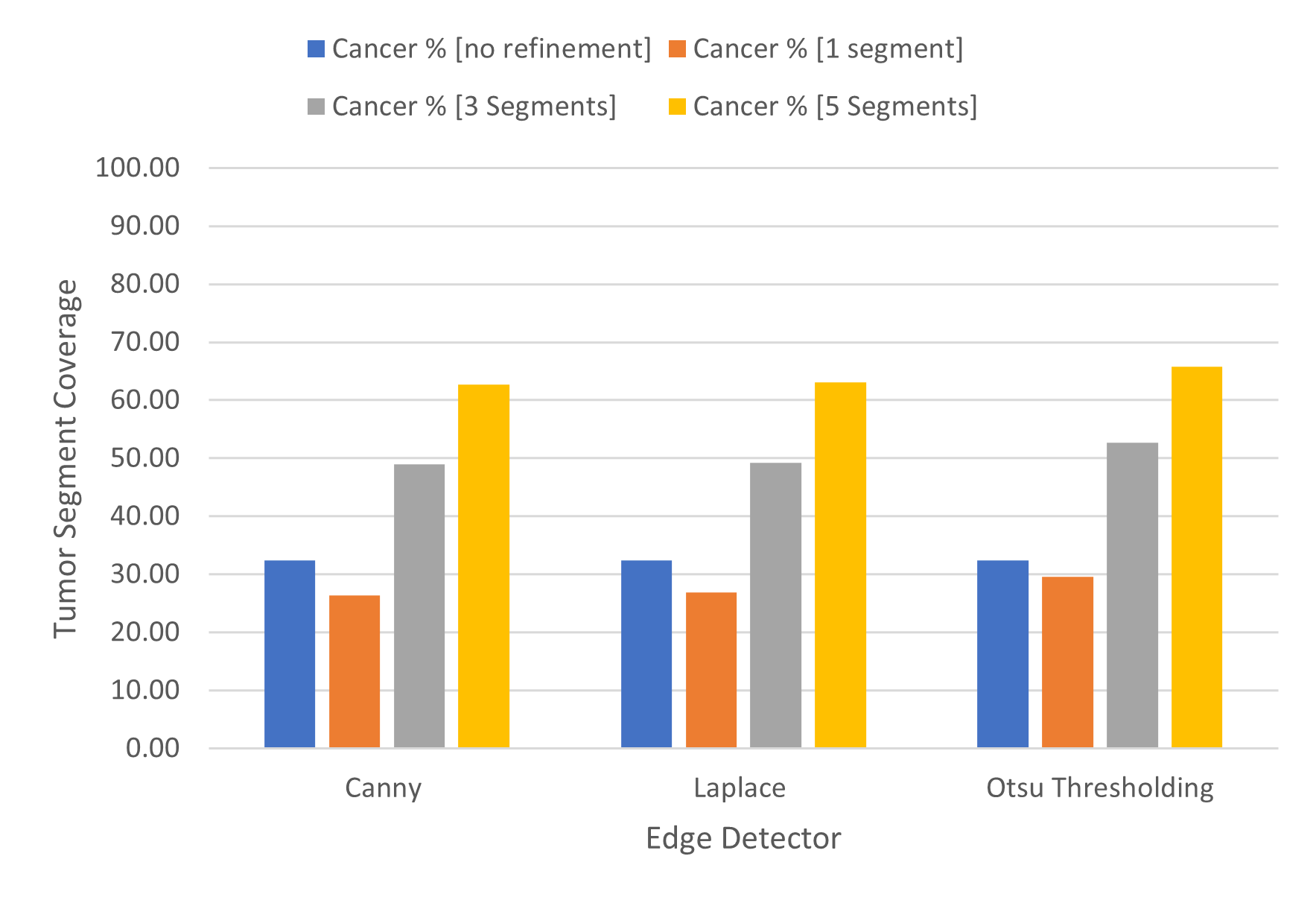}%
 \caption{Average Tumor Segment Coverage for each edge detector regarding the number of segments.}
\end{figure}

In order to check the balance between coverage and specificity, we used the brain segment coverage, where 1 segment covers 11.02\%, 3 segments cover 26.49\% and 5 segments covers 39.03\% on average, across the 3 techniques. Taking this into account, using 5 segments demonstrates an excessive number of non-informative segments, while using the 1 best segment results in limited tumor segment coverage and less accurate explanations. Given this consideration, the use of 3 segments emerges as the best choice, finding a balance between avoiding overuse of non-informative brain regions, and offering insightful explanations.

\begin{figure}[h]
    \centering
    \includegraphics[width=0.7\textwidth]{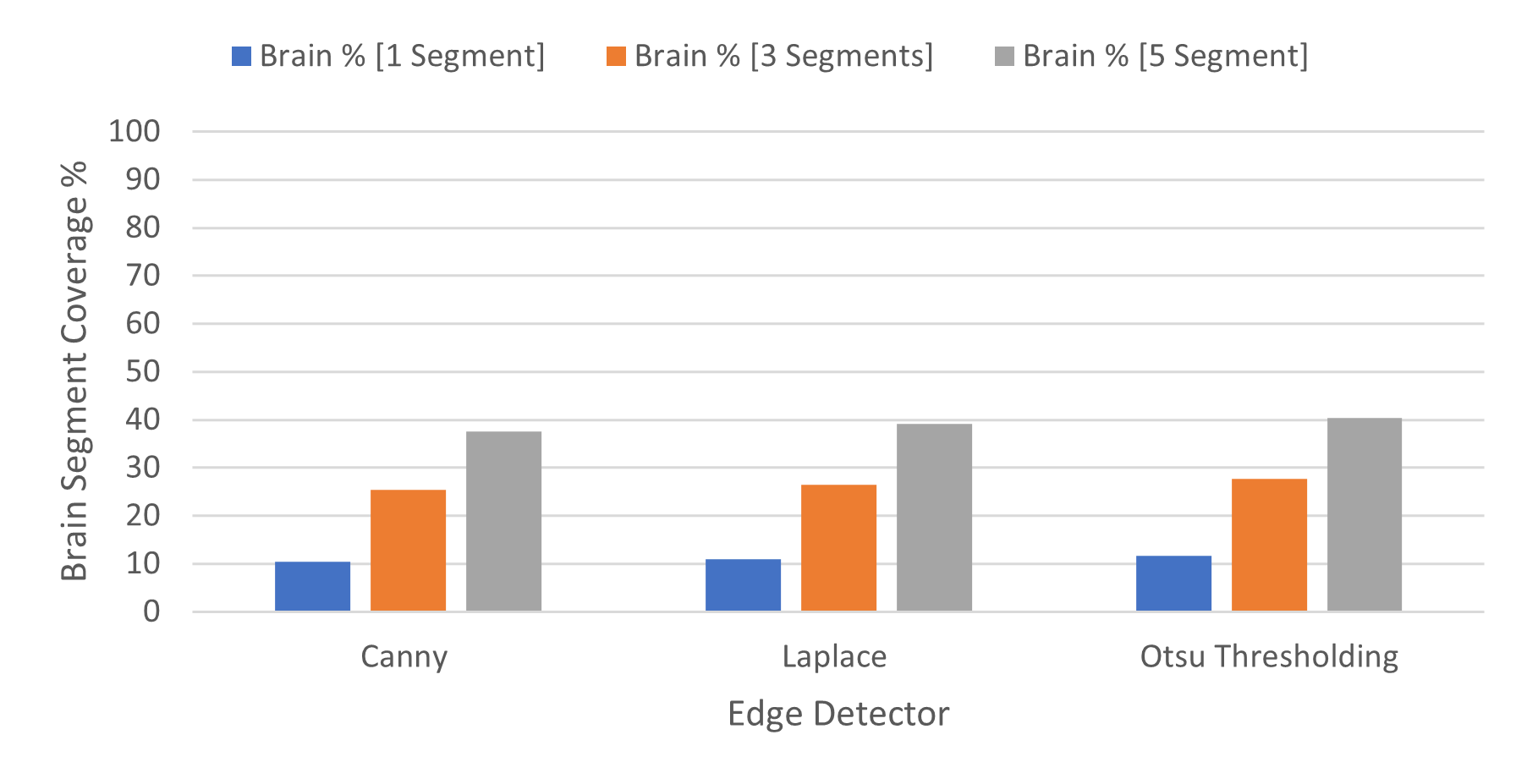}%
 \caption{Brain Segment Coverage Distributions for each edge detector regarding the number of segments.}

\end{figure}

\subsection{Advantages and Limitations}
\label{subsec:limitations}

While the proposed refinement mechanism has shown improvement in explainability, certain limitations should be aknowledged. One notable drawback observed in the proposed refinement mechanism is the potential inconsistency in creating the brain mask. The method relies on the edges detected by some edge detectors and subsequently extracting the largest contour as the brain mask. However, in some instances, this process may yield inconsistent results, generating a brain mask that covers the entire image, results in a completely blank mask, or produces a blank interior. Such variations in the brain mask could lead to misleading explanations, emphasizing the importance of a fine-tuned refinement approach.
\begin{figure}[h]
    \centering
    \includegraphics[width=0.5\textwidth]{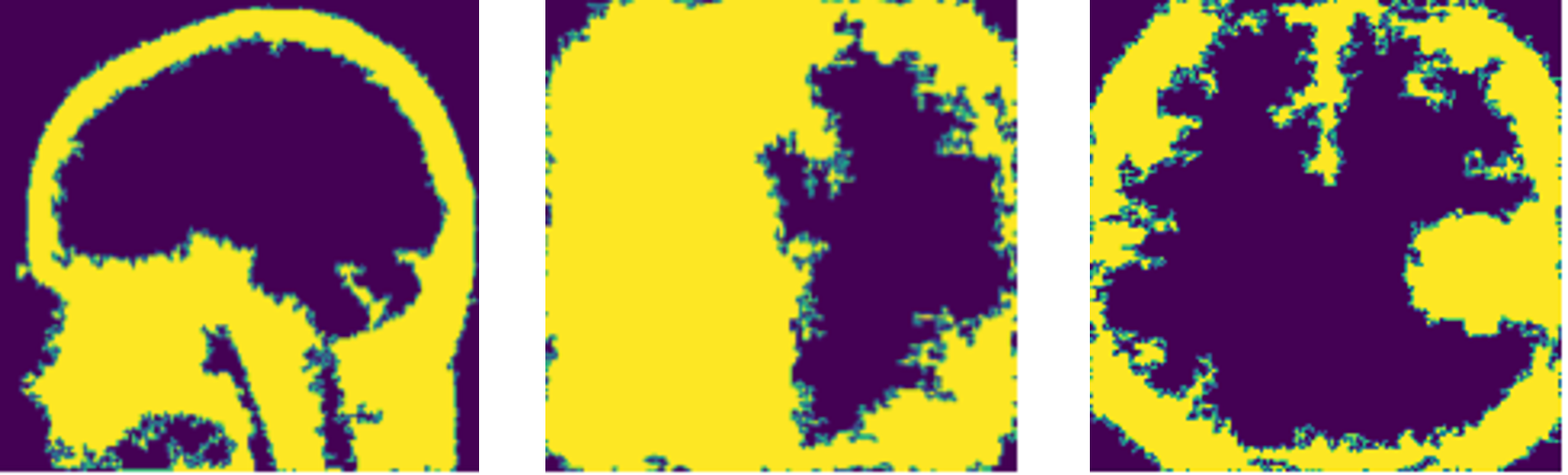}%
 \caption{Instances of wrong brain mask production.}
    \label{fig:enter-label}
\end{figure}

\section{Conclusions}
\label{sec:concl}

This work explores the topic of deep learning explainability in relation to the interpretation of brain tumor images. When combined with a new refining method, the LimeImageExplainer has shown promising progress in providing explanations for model predictions that are both more interpretable and accurate. Our suggested adjustment adds to the field's ongoing attempts to close the gap between complex model outputs and human interpretability by improving the specificity of segmented regions and addressing the weaknesses of initial explanations.

Although the results demonstrate the effectiveness of the refining mechanism, it is important to recognize the limits that have been noted, especially with regard to the consistency of brain mask production. To ensure reliable and effective brain mask production, further work should be focused on improving the refinement, possibly integrating dynamic thresholds for edge detection.

Overall, this work is a positive step toward improving the interpretability and transparency of deep learning models used in medical image analysis. The development of trustworthiness and the ease of integrating these models into clinical decision-making processes will depend on continuous efforts to improve explainability mechanisms as the field progresses.

\section{Aknowledgements}
This paper is a result of research conducted within the “MSc in Artificial
Intelligence and Data Analytics” of the Department of Applied Informatics of
University of Macedonia. The publication of the paper is funded
by the University of Macedonia Research Committee

\bibliographystyle{unsrtnat}
\bibliography{main}  

\end{document}